\documentclass[aps,prd,preprint,nofootinbib,superscriptaddress]{revtex4-1}
\usepackage{hyperref}
\usepackage{url}
\usepackage{color}
\usepackage{graphicx}
\usepackage{epsfig}
\usepackage{amsmath}
\def\beq{\begin{equation}}
\def\eeq{\end{equation}}
\def\beqa{\begin{eqnarray}}
\def\eeqa{\end{eqnarray}}
\def\lsim{\:\raisebox{-0.5ex}{$\stackrel{\textstyle<}{\sim}$}\:}

\def\half{\frac{1}{2}}

\def\erg{\,{\rm erg}}

\def\keV{\,{\rm keV}}
\def\MeV{\,{\rm MeV}}

\def\msun{M_\odot}
\def\eminus{e^{-}}

\def\nue{\nu_e}
\def\anue{\bar{\nu}_e}
\def\numu{\nu_\mu}
\def\anumu{\bar{\nu}_\mu}
\def\nutau{\nu_\tau}
\def\anutau{\bar{\nu}_\tau}
\def\nui{\nu_i}
\def\nux{\nu_x}
\def\anux{\bar{\nu}_x}
\def\Enu{E_\nu}
\def\Enumin{E_{\nu,{\rm min}}}
\def\Temax{T_{e,{\rm max}}}
\def\Enui{E_{\nui}}

\def\Enuiav{\langle\Enui\rangle}
\def\Enuisqav{\langle E^2_{\nui}\rangle}
\def\alphanui{\alpha_{\nui}}
\def\fznue{f^0_{\nue}}
\def\fzanue{f^0_{\anue}}
\def\fznumu{f^0_{\numu}}
\def\fzanumu{f^0_{\anumu}}
\def\fznutau{f^0_{\nutau}}
\def\fzanutau{f^0_{\anutau}}
\def\fznux{f^0_{\nux}}
\def\fnue{f_{\nue}}
\def\fanue{f_{\anue}}
\def\fnumu{f_{\numu}}
\def\fanumu{f_{\anumu}}
\def\fnutau{f_{\nutau}}
\def\fanutau{f_{\anutau}}
\def\GFsq{G_F^2}
\def\Qminus{Q_{-}}
\def\Qplus{Q_{+}}
\def\Qtminus{\tilde{Q}_{-}}
\def\Qtplus{\tilde{Q}_{+}}
\def\Enuiav{\langle\Enui\rangle}
\def\Enuisqav{\langle E^2_{\nui}\rangle}
\def\alphanui{\alpha_{\nui}}
\def\cenns{CE$\nu$NS}

\def\Xe132{^{132}{\rm Xe}}

\def\Cs132{^{132}{\rm Cs}}

\begin{document}
\title{Elastic scattering of supernova neutrinos with electrons 
in xenon}
\author{Pijushpani Bhattacharjee}
\email{pijushpaniw@gmail.com}
\affiliation{Department of Physics, Syracuse University, Syracuse NY
13244, USA}
\author{Kamales Kar}
\email{kamales.kar@gmail.com}
\affiliation{Ramakrishna Mission Vivekananda Educational and Research
Institute, Belur Math, Howrah 711202, India}

\begin{abstract}
Neutrinos from a Galactic core collapse supernova can undergo elastic 
scattering with electrons in xenon atoms in liquid xenon based dark 
matter detectors giving rise to electrons of kinetic energy up to 
a few MeV. We calculate the scattered electron spectrum and the number 
of such elastic scattering events expected for a typical Galactic core 
collapse supernova in a xenon target. Although the expected number of 
events is small (compared to, for example, inelastic neutrino-nucleus 
charged current interaction with xenon nuclei, that also gives rise to 
final state electrons), the distinct spectral shape 
of the scattered electrons may allow identification of the elastic 
scattering events. Further, while the process is dominated 
by neutrinos and antineutrinos of electron flavor, it receives 
contributions from all the six neutrino species. Identification of the  
electron scattering events may, therefore, allow an estimation of  
the relative fractions of the total supernova explosion energy going 
into electron flavored and non-electron flavored neutrinos. 
\end{abstract}
\maketitle

\section{Introduction}
\label{sec:intro}
Emission of bursts of neutrinos (and antineutrinos) of all flavors
with a total energy of $\sim 10^{53}\erg$ over a timescale of $\sim$ 10 
s and with average neutrino energies of $\sim$ 10 MeV is a robust 
prediction of the theory and phenomenology of core collapse supernova 
(CCSN) explosions; see, e.g., 
~\cite{ccsn-revs,Mirizzi-Sovan-rev_16,Burrows-papers}. 
This prediction was spectacularly confirmed by the detection of 
neutrinos from the supernova SN1987A located in the Large Magellanic 
Cloud (LMC) at a distance of $\sim$ 50 kpc from 
Earth~\cite{SN1987A-detection}. A number of currently operating and 
upcoming large neutrino detectors~\cite{SN-nu-detectors} should be able 
to detect neutrino bursts from reasonably close by (occurring within, 
say, 10 kpc from Earth) future Galactic supernova (SN) bursts. 

Of particular interest for the purpose of this paper is the possibility 
of SN neutrino detection in the next generation large liquid xenon based 
dark matter detectors~\cite{SN-nu-in-xenon,xenon-detectors-whitepaper}. 
The primary SN neutrino detection channel in these detectors is the weak 
neutral current (NC) process of coherent elastic neutrino-nucleus 
scattering (\cenns)~\cite{cenns} in which neutrinos and 
antineutrinos of all flavors elastically scatter off the 
xenon nuclei. The recoiling xenon nuclei give rise to scintillation  
photon signals as they lose energy in the liquid xenon medium. Because 
of the flavor blind nature of the \cenns\ process, identification of SN 
neutrino induced \cenns\  events in these detectors can in principle 
provide an 
estimate of the total SN explosion energy going into neutrinos. In 
addition to \cenns, SN $\nue$s can also undergo inelastic charged 
current (CC) interaction with the xenon nuclei~\cite{BBCGKS_22} 
producing a prompt electron in the final state together with gamma rays 
and neutrons from de-excitation of the target xenon nucleus. Being 
dominantly sensitive to 
$\nue$s~\cite{BBCGKS_22} and completely insensitive, due to energetic 
reasons, to neutrinos and antineutrinos of muon and tau flavors, the 
detection and identification of the electron events due to inelastic 
$\nue$-nucleus CC interaction may provide a 
handle on the SN explosion energy going into $\nue$s. 

The subject of this paper is a possible third kind of SN neutrino 
interaction process in xenon based detectors, namely, elastic 
scattering of the SN neutrinos with the electrons of the xenon atoms. As 
we shall see, due to 
relatively small cross section of the process, the expected number of 
elastic scattering interactions is small compared to, for example, 
inelastic neutrino-nucleus CC interactions~\cite{BBCGKS_22} that also 
produce an electron in the final state. However, the scattered electrons 
are found to have a distinct spectral shape that may allow 
distinguishing the elastic scattering events from the inelastic 
neutrino-nucleus CC origin electron events. In addition, although the 
elastic scattering process is dominated by neutrinos and antineutrinos 
of electron flavor, it receives contributions from all the six neutrino 
species. Identification of the electron scattering events may, 
therefore, allow an independent estimation of the relative fractions  
of the total supernova explosion energy going into electron flavored and 
non-electron flavored neutrinos.   

The rest of the paper is organized as follows: In section 
\ref{sec:nu-e-scatt-xsecs} we write down the cross 
sections for elastic scattering of all the neutrino and antineutrino 
species with the electrons bound within xenon atoms adopting the 
so-called ``free" electron approximation. In section 
\ref{sec:event-counts} we write down the expression for 
calculating the energy spectrum of the scattered electrons, 
i.e., differential number of scattered electrons as a function of 
their kinetic energy for a given SN neutrino flux spectrum. The SN 
neutrino flux spectrum we adopt for illustrating our calculations is 
described in section \ref{sec:SNnu-flux}. Section \ref{sec:Results} 
describes our results, and a brief summary of our main results 
is presented in section \ref{sec:Summary}. 

\section{Elastic scattering of neutrinos and antineutrinos on electrons 
in xenon}
\label{sec:nu-e-scatt-xsecs}
Since SN neutrinos have energies only up to a few tens of MeV, muon or 
tau lepton production in the final state is not possible. Therefore, we 
only consider the case of the final state charged 
lepton to be an electron. The differential cross section for elastic 
scattering of a $\nue$ off a free electron ($\nue\eminus \to 
\nue\eminus$), a process that occurs via both charged current (CC) as 
well as neutral current (NC) electroweak interactions, can be written 
as~\cite{tHooft_71,Marciano-Parsa_03,Gaspert-etal_22} 
\beq
\frac{d\sigma(\nue\eminus \to \nue\eminus)}{dT_e} = \frac{2\GFsq 
m_e}{\pi}\left[\Qminus^2 + \Qplus^2 \left(1-\frac{T_e}{\Enu}\right)^2 
- \Qminus\Qplus \frac{m_e 
T_e}{\Enu^2}\right]\Theta\left(\Enu-\Enumin\right)\,,
\label{eq:nue-e-diff-xsec}  
\eeq
where $\Enu$ and $T_e$ are the incident neutrino 
energy and the kinetic energy of the final state electron, respectively,  
$G_F=1.166\times10^{-11} \MeV^{-2}$ is the Fermi constant, $m_e$ is the 
electron rest mass,  
\beq
\Qminus=\half + \sin^2\theta_W\,, \,\,\,\, 
\Qplus=\sin^2\theta_W\simeq0.231\,,   
\label{eq:Qminus_Qplus} 
\eeq
and  
\beq
\Enumin=\half\left[T_e+\sqrt{T_e(T_e+2m_e)}\right]
\label{eq:Enumin}
\eeq
is the minimum neutrino energy that can give rise to a recoiling 
electron of kinetic energy $T_e$. 

For the process $\nux\eminus \to \nux\eminus$ 
(with $\nux=\numu\,, \nutau$), which is a purely NC process, the cross 
section is given by~\cite{Marciano-Parsa_03}  
\beq
\frac{d\sigma(\nux\eminus \to \nux\eminus)}{dT_e} = \frac{2\GFsq
m_e}{\pi}\left[\Qtminus^2 + \Qtplus^2 
\left(1-\frac{T_e}{\Enu}\right)^2 -\Qtminus\Qtplus \frac{m_e 
T_e}{\Enu^2}\right]\Theta\left(\Enu-\Enumin\right)\,,
\label{eq:nux-e-diff-xsec}
\eeq
where 
\beq
\Qtminus=-\half + \sin^2\theta_W\,, \,\,\,\, {\rm and} \,\,\,\,
\Qtplus=\sin^2\theta_W\,.
\label{eq:Qtminus_Qtplus}
\eeq
The cross sections for the antineutrino species, $\anue\eminus \to 
\anue\eminus$ and $\anux\eminus \to \anux\eminus$, are given by   
equations (\ref{eq:nue-e-diff-xsec}) and (\ref{eq:nux-e-diff-xsec}), 
respectively, with $\Qminus$ ($\Qtminus$) and $\Qplus$ ($\Qtplus$) 
interchanged. 

\begin{figure}[htb]
\centering
\includegraphics[width=0.9\columnwidth]{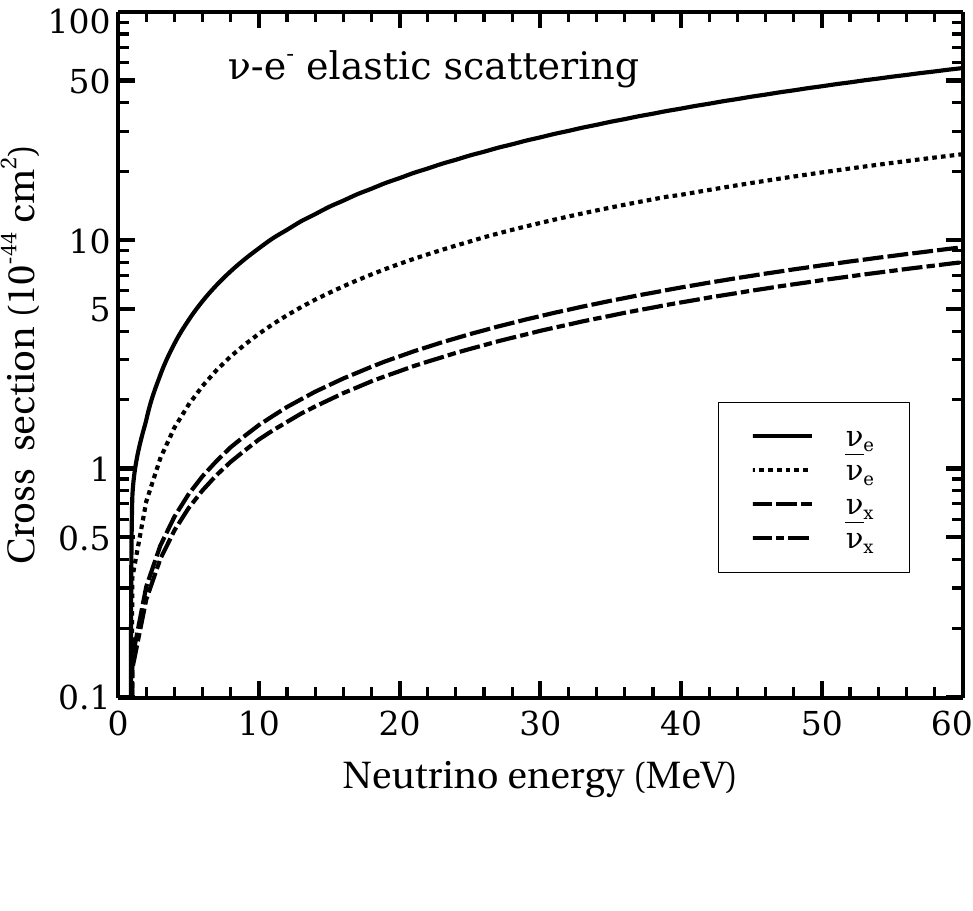}
\vskip-1cm
\caption{The total neutrino-electron elastic scattering cross sections 
as functions of neutrino energy for different neutrino species with 
$\nux \equiv (\numu, \nutau)$.}
\label{fig:total-xsecs}
\end{figure}

The total cross sections for a given initial neutrino energy can be 
obtained by simply integrating the differential cross sections 
given above from $T_e=0$ to $\Temax=\Enu/(1+\frac{m_e}{2\Enu})$ (given 
by eq.~(\ref{eq:Enumin})). The total elastic scattering cross sections 
as functions of neutrino energy for different neutrino species are shown 
in Fig.~\ref{fig:total-xsecs}. 

For neutrinos scattering off electrons in xenon atoms (atomic 
number $Z=54$) we use, for simplicity, the ``free" electron (FE) 
approximation whereby the atomic electrons are assumed to be free and 
at rest. This should be a reasonably good 
approximation~\cite{Gaspert-etal_22,Voloshin_10_11} for 
neutrino 
energies and final state electron kinetic energies satisfying $\Enu\gg 
{\rm max}(T_e, E_b)$, where $E_b$ is the largest relevant atomic binding 
energy, which 
for xenon is $\sim 35\keV$. The SN neutrinos have average energies of 
$\sim 10\MeV$, and the neutrino fluxes also peak at, and fall  
off below and above, energies of $\sim 
10\MeV$~\cite{Mirizzi-Sovan-rev_16}. For SN neutrinos, 
therefore, the above condition for applicability of the FE approximation 
is well satisfied for $T_e \lsim$ few hundred keV, the energy 
region where, as we shall see below (see 
Fig.~\ref{fig:elastic-scatt_vs_CC_comp}), the dominant contribution 
to the scattered electron spectrum lies in the case of xenon.       

In the FE approximation, then, the differential cross section for 
elastic scattering of neutrinos off electrons in xenon atoms is simply 
given by 
\beq
\frac{d\sigma^{\rm Xe}(\nui\eminus \to \nui\eminus)}{dT_e} = 
54\, \frac{d\sigma(\nui\eminus \to \nui\eminus)}{dT_e}\,,
\label{eq:nui-e-Xe-diff-xsec}
\eeq
where $\nui\equiv (\nue, \anue, \numu, \anumu, \nutau, \anutau$), with 
the individual free electron cross sections as given above.  

\section{Number of neutrino-electron elastic scattering events in xenon  
due to a supernova neutrino burst}
\label{sec:event-counts} 
With the cross sections specified as above, the differential number of 
$\nu-\eminus$ elastic scattering events in xenon per ton (1000 kg) of 
xenon as a function of the scattered electron kinetic energy $T_e$ can 
be written as     
\beq
\frac{dN}{dT_e} = n_{\rm Xe}\sum_{\nui}\int d\Enu \Phi_{\nui}(\Enu)
\frac{d\sigma^{\rm Xe}(\nui\eminus \to \nui\eminus)}{dT_e}\,, 
\label{event-spect}
\eeq
where $\nui\equiv (\nue, \anue, \numu, \anumu, \nutau, 
\anutau)$, $\Phi_{\nui}(\Enu)$ is the time-integrated (over 
the SN burst duration) flux (number 
per unit area per unit energy) at Earth of the $\nui$ species, 
and $n_{\rm Xe}$ is the number of xenon atoms per ton of xenon. For the 
purpose of illustration of our results here we consider a 
specific isotope of xenon, namely, $\Xe132$, for which $n_{\rm 
Xe} = 4.56\times10^{27}$. 

\section{Supernova neutrino flux}
\label{sec:SNnu-flux}
The time dependent number of neutrinos of type $\nui$ 
emitted (at the neutrinosphere) per unit time per unit energy is usually 
written  as~\cite{Keil-etal_03}
\beq
f^0_{\nui} (t,\Enu) = \frac{L_{\nui} (t)}{\Enuiav 
(t)}\varphi_{\nui}(\Enu,t)\,,
\label{eq:fnuzero}
\eeq
where $L_{\nui} (t)$, $\Enuiav (t)$ and $\varphi_{\nui}(\Enu,t)$ are, 
respectively, the luminosity, average energy and 
normalized energy spectrum ($\int 
\varphi_{\nui}(\Enu,t)d\Enu = 1$) of neutrinos of type $\nui$. The 
energy spectrum is parametrized as~\cite{Keil-etal_03} 
\beqa 
\displaystyle{\varphi_{\nui}(\Enu,t)} & = \displaystyle{\frac{1}{\Enuiav 
(t)}}\displaystyle{\frac{\bigl(1+\alphanui(t)\bigr)^{1 
+\alphanui(t)}}{\Gamma\bigl(1+\alphanui (t)\bigr)}}
\displaystyle{\left(\frac{\Enu}{\Enuiav(t)}\right)^{\alphanui(t)}}\,
\nonumber\\
 & \,\,\,\, \times\,\, 
\displaystyle{\exp\left[-\bigl(1+\alphanui 
(t)\bigr)\frac{\Enu}{\Enuiav(t)}\right]}\,, 
\label{eq:phi}
\eeqa
where
\beqa
\alphanui (t)=\frac{2\Enuiav^2(t) - \Enuisqav(t)}{\Enuisqav (t) - 
\Enuiav^2 (t)}\nonumber
\eeqa 
is the spectral shape parameter. 

For values of the quantities $L_{\nui}(t)$, $\Enuiav(t)$ and 
$\alphanui(t)$, in this paper we use for definiteness the results from 
the hydrodynamic simulations of the Basel-Darmstadt (B-D) 
group~\cite{Basel-Darmstadt_10} for a $18\msun$ progenitor star. We note 
here that $\numu$, $\anumu$, $\nutau$, and $\anutau$ have the same 
luminosities, average energies and spectral shape parameters in the B-D 
simulations. 

The fluxes of neutrinos of different flavors arriving at Earth are 
subject to various flavor oscillation processes including the 
conventional matter-enhanced Mikheyev-Smirnov-Wolfenstein (MSW) 
oscillation (see, e.g., \cite{Mohapatra-Pal-book_04}) as well as 
collective 
neutrino flavor oscillations due to $\nu$-$\nu$ interaction in the  
interior of the SN progenitor 
star~\cite{Mirizzi-Sovan-rev_16,Duan-etal_10,sovan-etal-NPB_16,Capozzi-Dasgupta-etal_19}. 
The phenomenon of collective oscillation can be complicated 
depending on the spectral details 
of different neutrino species and matter density profile inside the SN 
progenitor star. Here we shall follow a simplified 
treatment~\cite{Mirizzi-Sovan-rev_16,Borriello-Sovan-etal_12} in which, 
for the time-integrated fluxes of our interest here, the collective 
oscillation effects are generally regarded as  
small and the fluxes, $\Phi_{\nui} (\Enu)\equiv \frac{1}{4\pi d^2}\int 
dt f_{\nui} (t,\Enu)$, of various neutrino species arriving at Earth 
($d$ being the distance to the SN from Earth) are 
essentially determined by the standard MSW oscillations (with 
$\theta_{13} = 0$, $\theta_{23} = 45^{\circ}$, 
$\theta_{12} = 33^{\circ}$), and are 
given, for the case of normal hierarchy (NH) of neutrino mass ordering,  
by  
\beq
\fnue = \fznux \quad {\mathrm{and}}\quad \fanue =  
\cos^2\theta_{12}\fzanue + \sin^2\theta_{12}\fznux\,,
\label{eq:fnue_fanue}
\eeq
\beq
\fnumu = \fznue \quad {\mathrm{and}}\quad \fanumu =  
\cos^2\theta_{12}\fznux + \sin^2\theta_{12}\fzanue\,,
\label{eq:fnumu_fanumu}
\eeq
\beq
\fnutau = \fznux \quad {\mathrm{and}}\quad \fanutau =  
\fznux\,,
\label{eq:fnutau_fanutau}
\eeq
where as mentioned above we have used 
$\fznumu=\fzanumu=\fznutau=\fzanutau\equiv\fznux$ 
following the results of the hydrodynamic simulations of the 
Basel-Darmstadt group~\cite{Basel-Darmstadt_10} that we use for the 
emitted fluxes of neutrinos of different flavors in our numerical 
calculations described below. 

For the case of inverse hierarchy (IH) of neutrino mass ordering, the 
fluxes of various neutrino species are obtained by interchanging the 
neutrinos and corresponding antineutrinos of each flavor on both 
sides of equations (\ref{eq:fnue_fanue})--(\ref{eq:fnutau_fanutau}) and 
at the same time interchanging $\cos^2\theta_{12}$ and 
$\sin^2\theta_{12}$. 

With the fluxes of various neutrino species and their elastic 
scattering cross sections with electrons in xenon atoms specified as 
above, we 
can easily evaluate the expected number of scattering events as a 
function of the scattered electron's kinetic energy using equation 
(\ref{event-spect}). The results are shown and discussed below. We show 
the results only for the NH case of neutrino mass ordering; 
there are no significant differences in the results for the IH case.  
Also, for convenience, we show the results with the SN at a distance of 
$d=1$ kpc from Earth, with the number of events simply scaling as 
$1/d^2$. 
\section{Results and discussions}
\label{sec:Results}
\begin{figure}[htb]
\centering
\includegraphics[width=0.9\columnwidth]{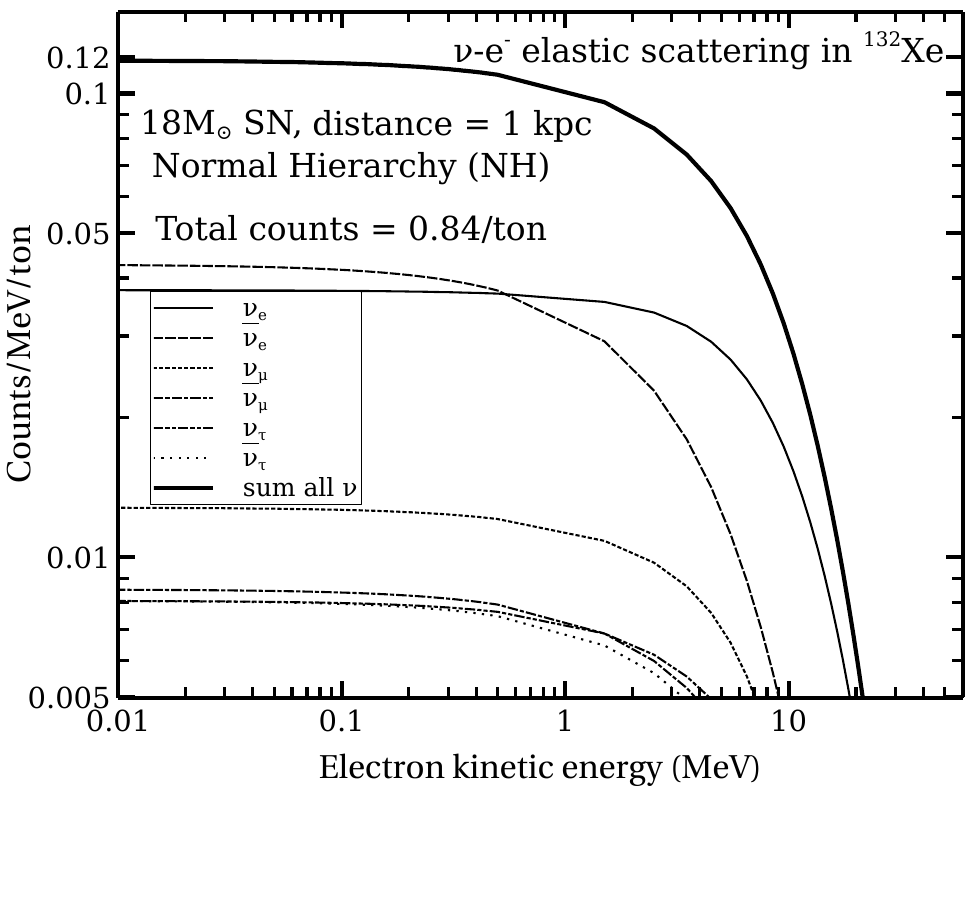} 
\vskip-1cm
\caption{The differential spectrum of scattered electrons due to 
supernova neutrino induced neutrino-electron elastic scattering 
on a target of $\Xe132$.} 
\label{fig:diff-spect-NH}
\end{figure}

The differential spectrum of the scattered electrons due to
supernova neutrino induced neutrino-electron elastic scattering
on a target of $\Xe132$ is shown in Figure \ref{fig:diff-spect-NH}. It 
is seen that contributions from $\nue$s and $\anue$s dominate the 
total spectrum: While $\nue$ and $\anue$ contributions are comparable, 
$\anue$ contribution slightly exceeds that of $\nue$ up to $T_e \sim$  
0.5 MeV above which the reverse is the case. The combined number of 
$\nue$ and $\anue$ induced events is typically a factor of 2 larger than 
the combined number of events due to $\numu$, $\anumu$, $\nutau$ and 
$\anutau$. At 
the same time, the $\numu$ contribution is about a factor of 1.5 larger 
than the individual $\anumu$, $\nutau$ and $\anutau$ contributions 
which are comparable with each other.   
\begin{figure}[htb]
\centering
\includegraphics[width=0.7\columnwidth]{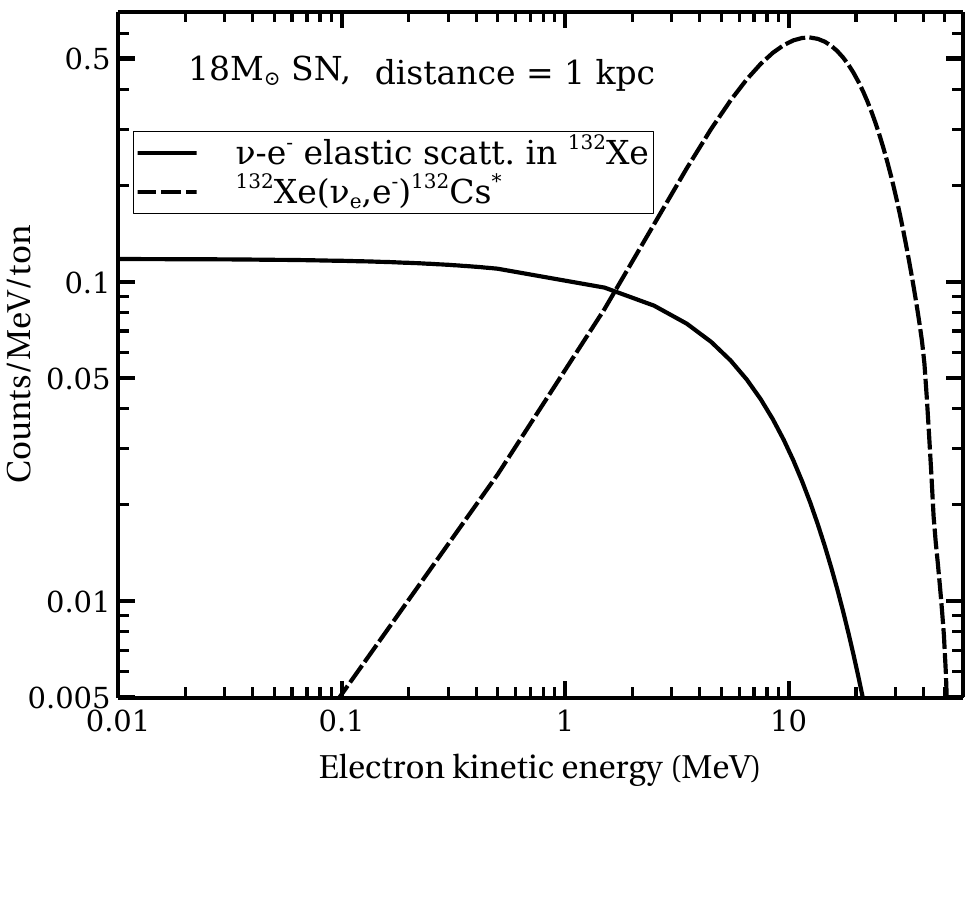}
\vskip-1cm
\caption{Comparison of the differential spectrum of (a) the scattered 
electrons due to neutrino-electron elastic 
scattering on a target of $\Xe132$ (solid line), and (b) the 
electrons due to $\nue$-induced inelastic neutrino-nucleus charged 
current interaction, $\Xe132 (\nue, 
\eminus)\Cs132^*$~\cite{BBCGKS_22}(dashed line), for 
the same SN model as used in Fig.~\ref{fig:diff-spect-NH}.} 
\label{fig:elastic-scatt_vs_CC_comp}
\end{figure}

In Fig.~\ref{fig:elastic-scatt_vs_CC_comp} we show for comparison  
the differential spectrum of the scattered electrons due to 
neutrino-electron elastic scattering on a target of $\Xe132$ and that 
of the electrons produced through $\nue$-induced inelastic 
neutrino-nucleus charged current interaction, $\Xe132 (\nue, 
\eminus)\Cs132^*$~\cite{BBCGKS_22}, both due 
to the same SN model as used in Fig.~\ref{fig:diff-spect-NH}. 
We see that although in 
terms of total number of events the $\nue$ induced 
inelastic neutrino-nucleus CC interaction dominates over the  
$\nu-\eminus$ elastic scattering process --- a total count of $\sim$ 
13/ton for the former as against 0.84/ton for the elastic scattering 
process --- the latter dominates for electron kinetic energies below 
$\sim$ 1 MeV. Identification of low energy ($\lsim$ 100 
keV, say) electron events may, therefore, offer a signature of the 
$\nu-\eminus$ elastic scattering process that receives contributions 
from all the six species of neutrinos with varying degrees of 
contribution to the total number of scattering events, as discussed 
above. Thus, together with (a) the nuclear recoil events due to the 
\cenns\  process, which could yield an estimate of the total SN 
explosion energy going into all six neutrino species, and (b) the $\nue$ 
induced inelastic neutrino-nucleus CC interactions, which would provide 
an estimate of the energy fraction going into $\nue$s, the 
identification of the $\nu-\eminus$ elastic scattering events may offer 
a handle on the relative fractions of the total supernova explosion 
energy going into different neutrino species, in particular, the 
relative energy fractions going into electron flavored and non-electron 
flavored neutrinos. 

\section{Summary} \label{sec:Summary} In this paper we have studied the 
process of elastic scattering of supernova neutrinos on electrons in 
xenon atoms as a possible detection channel for supernova neutrinos in 
the next generation multi-ton scale liquid xenon based dark matter 
detectors. This detection channel would be in addition to the channels 
offered by \cenns- and $\nue$-nucleus inelastic charged current 
interaction processes. We find that while the expected number of 
$\nu-\eminus$ elastic scattering events is small compared to the other 
final-state electron producing process, namely, inelastic $\nue$-nucleus 
charged current interaction with xenon nuclei, the distinct spectral 
shape of the scattered electrons may allow identification of the elastic 
scattering events. Together with \cenns-induced nuclear recoil events 
and the electron events due to inelastic $\nue$-nucleus charged current 
interaction, the identification of the $\nu-\eminus$ elastic scattering 
events may offer a handle on the total supernova explosion energy going 
into neutrinos of all species and at the same time the relative 
fractions of the total supernova explosion energy that go into electron 
flavored and non-electron flavored neutrinos. For a realistic assessment 
of the feasibility of detection and identification of the elastic 
scattering events one needs to consider the spectrum of the 
scintillation light yield due to the recoil electrons as well as all the 
relevant backgrounds in the liquid xenon detectors of interest. While 
the transient nature of the burst may be a mitigating factor as far as 
the background is concerned, this remains to be examined in detail. 
Further, it will be of interest to carry out detailed calculations 
of the event rates using supernova neutrino fluxes from recent 
2D and 3D numerical simulations of core-collapse supernova explosions 
(e.g., \cite{Burrows-papers}). We hope to consider these issues in a 
future work.

\section*{Acknowledgments}
We thank Sovan Chakraborty for providing the numerical data used in 
this paper for the fluxes of different neutrino species from the results 
of the Basel-Darmstadt supernova simulations~\cite{Basel-Darmstadt_10}. 
One of us (PB) wishes to thank Jennifer Ross for hosting him as a 
Visiting Scholar in the Physics Department, Syracuse University. 


\end{document}